\documentclass[reprint, twoside, 10pt, superscriptaddress, amsmath, amssymb, aps, prl, colorlinks = true,linkcolor = blue, citecolor = blue]{revtex4-2}

\usepackage{braket}
\usepackage{graphicx}
\usepackage{dcolumn}
\usepackage{bm}
\usepackage{mathrsfs}
\usepackage[ruled,vlined]{algorithm2e}
\usepackage{orcidlink}
\usepackage{amsmath}
\usepackage{hyperref}
\usepackage{cleveref}
\usepackage{appendix} 

\begin{document}

\title{Practical robust Bayesian spin-squeezing-enhanced quantum sensing under noises}

\def\QPMSZU{Institute of Quantum Precision Measurement, State Key Laboratory of Radio Frequency Heterogeneous Integration,\\ College of Physics and Optoelectronic Engineering, Shenzhen University, Shenzhen 518060, China}

\def\SYSUZH{Laboratory of Quantum Engineering and Quantum Metrology, School of Physics and Astronomy,\\ 
Sun Yat-Sen University (Zhuhai Campus), Zhuhai 519082, China}

\def\GD{Quantum Science Center of Guangdong-Hong Kong-Macao Greater Bay Area (Guangdong), Shenzhen 518045, China}

\author{Jinye Wei}
  \affiliation{\QPMSZU}
  \affiliation{\SYSUZH}

\author{Jungeng Zhou}
  \affiliation{\QPMSZU}
  \affiliation{\SYSUZH}

\author{Yi Shen}
  \affiliation{\QPMSZU}
  \affiliation{\SYSUZH}

\author{Jiahao Huang~\orcidlink{0000-0001-7288-9724}}%
  \email{Email: hjiahao@mail2.sysu.edu.cn, eqjiahao@gmail.com}
  \affiliation{\QPMSZU}
  \affiliation{\SYSUZH}  

\author{Chaohong Lee~\orcidlink{0000-0001-9883-5900}}%
  \email{Email: chleecn@szu.edu.cn, chleecn@gmail.com}
  \affiliation{\QPMSZU}
  \affiliation{\GD}

\date{\today}
\begin{abstract}
Spin-squeezed states constitute a valuable entanglement resource capable of surpassing the standard quantum limit (SQL). 
However, spin-squeezed states only enable sub-SQL uncertainty within a narrow parametric window near some specific points.
Identifying optimal measurement protocols for spin-squeezed states remains an outstanding challenge.
Here we present an adaptive Bayesian quantum estimation protocol that achieves optimal measurement precision with spin-squeezed states under noises. 
Our protocol operates by maintaining measurements near the optimal point and employing Bayesian inference to sequentially perform phase estimation, enabling robust high-precision measurement.
To account for realistic experimental conditions, we explicitly incorporate phase noises into the Bayesian likelihood function for more accurate estimation. 
Our protocol can be applied to various scenarios, such as quantum gravimeters and atomic clocks, achieving precision enhancement over conventional fitting methods under noises.
Our approach offers superior precision and enhanced robustness against noises, making it highly promising for squeezing-enhanced quantum sensing.

\end{abstract}

 
\maketitle

\textit{Introduction. --} 
Multiparticle entanglement is a crucial resource for enhancing quantum metrology beyond the standard quantum limit (SQL)~\cite{PhysRevLett.96.010401,PhysRevLett.97.150402,escher2011general,giovannetti2011advances,RevModPhys.90.035005,10.1063/5.0204102}. 
%
%
%
%
%
Spin-squeezed states, as a paradigmatic class of entangled states, have been extensively used to reduce measurement fluctuations~\cite{esteve2008squeezing,gross2010nonlinear,riedel2010atom,pedrozo2020entanglement,RN17,eckner2023realizing} and are far more robust against decoherence than Greenberger-Horne-Zeilinger states~\cite{10.1063/5.0204102}.
Generally, the minimal phase fluctuation for an $N$-particle spin-squeezed state is given by $\Delta \phi = {\xi/\sqrt{N}}$ with the spin squeezing parameter $\xi<1$~\cite{PhysRevA.47.5138,PhysRevA.46.R6797,PhysRevA.50.67,MA201189}.
However, the possible advantages offered by spin squeezing are still unclear to fully exploit.  
This is because spin-squeezed states can only reduce the phase uncertainty $\Delta \phi$ below the SQL (i.e. $\Delta \phi < {1/\sqrt N}$) when the phase $\phi$ is sufficiently close to some specific optimal values~\cite{PhysRevA.46.R6797,PhysRevLett.125.210503,gross2010nonlinear,RN18}.
Otherwise, $\Delta \phi$ will rapidly degrade when $\phi$ is away from the optimal point. 
In particular, the range of $\Delta \phi < {1/\sqrt N}$ decreases with $\xi$, that is, stronger squeezing corresponds to narrower range of sub-SQL precision.
Unfortunately, since $\phi$ is an unknown parameter to be measured, it is usually far away from the optimal point.
Moreover, under realistic experimental conditions, the squeezing-enhanced sensitivity is further constrained by noises. 
These limitations pose significant challenges for practical implementations of squeezing-enhanced quantum sensing.

\begin{figure*}[htbp]
\includegraphics[width=0.95\linewidth]{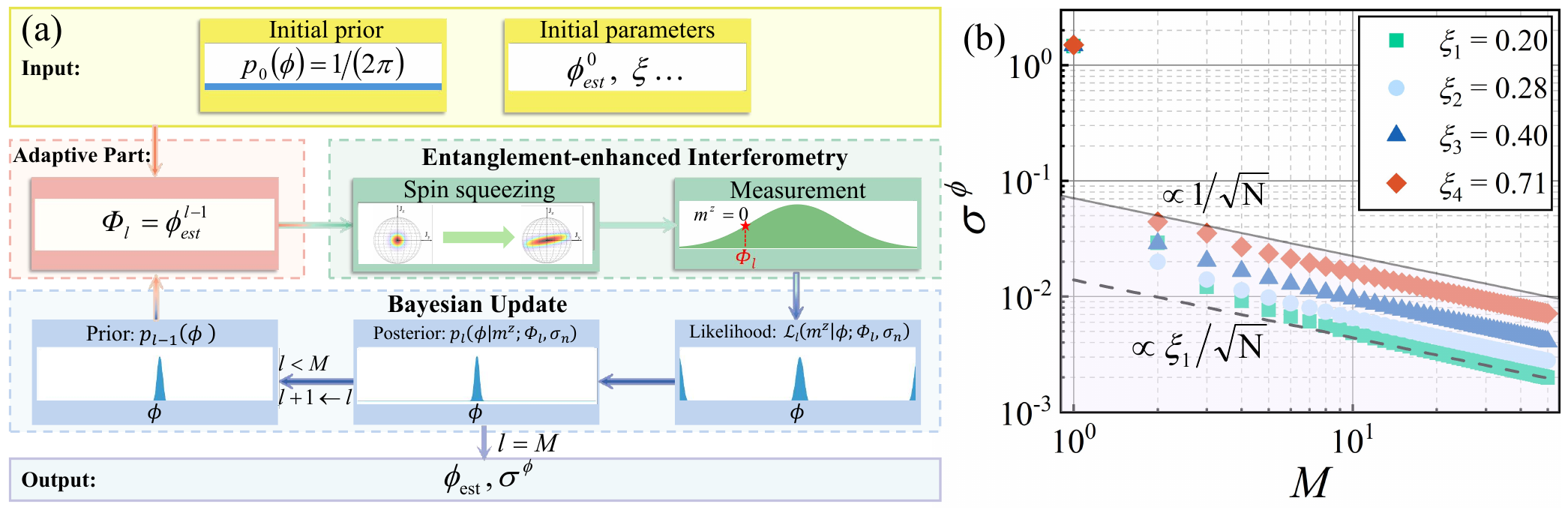}%
\caption{\label{BayesianPhaseEstimation} (color online).
(a) Schematic of adaptive Bayesian spin-squeezing-enhanced phase estimation. 
Given a prior distribution $p_{l-1}$ and the likelihood function, the posterior distribution $p_l$ is given by the Bayes' formula. 
The estimate $\phi_{est}^ {l}$ and its standard deviation $\sigma^{\phi}_l$ can be computed with $p_l$. 
Then, $\Phi_l$ is given to perform optimal measurement for each step. 
Generally, the initial prior $p_0$ can be chosen as an uniform distribution. 
(b) Phase precisions for different spin squeezing parameters $\xi$.}
\end{figure*} 
 
Bayesian quantum estimation (BQE) can combine a sequence of measurements to update probability distributions through Bayes’ rule. 
By combing measurements with different control parameters, BQE achieves both high precision and high dynamic range, which has been successfully applied in quantum phase estimation~\cite{PhysRevA.76.033613,PhysRevLett.117.010503,PhysRevLett.118.100503,wang2017experimental,PhysRevApplied.10.044033,RN4,Qiu_2022,gebhart2023learning,PhysRevA.109.042412,PhysRevA.109.062609}, quantum magnetometers~\cite{RN3,RN5,PhysRevX.7.031050,PhysRevX.9.021019,PhysRevApplied.16.024044,Craigie_2021,PhysRevA.106.052603,doi:10.1126/sciadv.adt3938}, atomic clocks~\cite{PhysRevApplied.22.044058,RN6,2024arXiv241114944Z} and atomic gravimeters~\cite{PhysRevResearch.7.L012064}.
Using Bayesian inference, BQE can adaptively optimize parameter estimation toward high precision by introducing an auxiliary parameter.
\textit{Can one introduce an auxiliary phase through BQE to adaptively push spin-squeezed-state-based metrology working on its optimal point?} 
In practice, the performance of BQE is severely affected by noises. 
Some preliminary studies have incorporated noise by using a reduced effective atom number in the likelihood function~\cite{PhysRevApplied.22.044058,doi:10.1126/sciadv.adt3938,10.21468/SciPostPhys.17.1.014}.
Choosing a spin component $\hat{J}_\alpha$ (with $\alpha\in\{x,y,z\}$) as the observable, the likelihood functions for spin-squeezed states can be expressed as Gaussian distributions~\cite{MA201189}.
However, it remains unclear how noise distorts these Gaussian functions in a BQE process.

In this Letter, we propose an adaptive Bayesian squeezing-enhanced protocol for quantum sensing to achieve robust and high-precision measurement under noises.
Our protocol uses adaptive BQE to adaptively lock spin-squeezed-state-based interferometry onto its optimal point, achieving sub-SQL precision. 
Simultaneously, it enhances robustness by reshaping the likelihood function to account for phase noises.
We demonstrate these capabilities in two key applications: quantum gravimetry and atomic clocks. 
In quantum gravimetry, the protocol maintains high precision and large dynamic range by combining a sequence of measurements with increasing interrogation times, overcoming limitations of conventional fringe-fitting. 
For atomic clocks, it significantly improves the Allan deviation under Bayesian locking, even in the presence of white, flicker, and random walk noises.
Our Bayesian estimation protocol is readily applied to enhance the performances of all existing spin-squeezed-state-based interferometry.

\textit{Adaptive Bayesian quantum phase estimation with spin-squeezed states. --}
We consider a system composed of $N$ two-mode particles, whose two modes are respectively denoted by $| a \rangle$ and $| b \rangle$.
The system can be well described by collective spin operators $\hat{J}_{\alpha}=\frac{1}{2}\sum_{l=1}^{N}\sigma_{\alpha}^{(l)}$ with $\alpha \in \{x, y, z\}$, where $\sigma_{\alpha}^{(l)}$ are the Pauli matrices for the $l$-th particle.
These collective spin operators satisfy angular momentum commutation relations: $\left[\hat{J}_{\alpha}, \hat{J}_{\beta}\right]=i\epsilon_{\alpha, \beta, \gamma}\hat{J}_{\gamma}$ with $\epsilon_{\alpha, \beta, \gamma}$ being the Levi-Civita symbol.
In the Schwinger representation, the collective spin operators can be expressed as: $\hat{J}_x=\frac{1}{2}(\hat{a}^\dagger\hat{b}+\hat{a}\hat{b}^\dagger)$, $\hat{J}_y=\frac{1}{2i}(\hat{a}^\dagger\hat{b}-\hat{a}\hat{b}^\dagger)$ and $\hat{J}_z=\frac{1}{2}(\hat{a}^\dagger\hat{a}-\hat{b}^\dagger\hat{b})$ with the annihilation operators $\hat{a}$ and $\hat{b}$ respectively for the particles in $| a \rangle$ and $| b \rangle$.

Without loss of generality, we analyze the phase estimation via Ramsey interferometry with an input spin-squeezed state $\ket{\psi}_{SS}=\frac{1}{\sqrt{\mathcal{N}_{S}}} \sum_{n=-N/2}^{N/2} e^{-\frac{n^2}{s^2 N}} \ket{n}_y$.
Here, $\hat J_y\ket{n}_y= n \ket{n}_y$ (with $n=-\frac{N}{2},...,\frac{N}{2}-1,\frac{N}{2}$) and $\mathcal{N}_{S}=\sum_{n=-N/2}^{N/2} e^{-\frac{2 n^2}{s^2 N}}$ (where $0<s<1$).
Through a phase accumulation $e^{-i\phi\hat J_z}$ and a recombination operation $e^{-i\frac{\pi}{2}\hat J_y}$, the output state becomes $\ket{\psi}_{out}=e^{-i\frac{\pi}{2}\hat J_y}e^{-i\phi\hat J_z}\ket{\psi}_{SS}$. 
The accumulated phase $\phi$ can be inferred by measuring the final half-population difference $\langle \hat J_z\rangle_{out}=\langle \hat J_x\rangle_{SS} \cdot \sin \phi \approx \mathcal{A} \sin \phi$ with $\mathcal{A}=\frac{N}{2} e^{-1/(2s^2N)}$ when $s^2 N > 1$. 
%
When $\phi\approx0$, according to the error propagation formula, we have the uncertainty
\begin{equation}\label{optimalpoint}
    \Delta \phi = \frac{\Delta \hat{J}_z}{\lvert{\partial\langle\hat{J}_z\rangle}/{\partial\phi}\rvert}
   =\frac{\sqrt{\langle\hat{J}_z^2\rangle-\langle \hat{J}_z \rangle^2}}{\lvert{\partial\langle\hat{J}_z\rangle}/{\partial\phi}\rvert}  \approx \frac{\xi}{\sqrt N}, 
\end{equation}
where the spin squeezing parameter $\xi=\sqrt{N} (\Delta\hat J_y)_{SS}/{\langle \hat J_x\rangle_{SS}}=s e^{1/(2 s^2 N)}$. 
However, the spin-squeezed states can only achieve sub-SQL precision within a narrow phase range near $\phi=0$. 
Otherwise the precision may even become worse than the SQL when the phase is far way from $\phi=0$. 
Thus, when employing conventional fringe-fitting across a broad phase range for measurement, the precision may be worse than the SQL even with spin-squeezed states~\cite{SM}.
Below, we overcome this problem by integrating adaptive BQE with spin-squeezed states, enabling robust sub-SQL measurements.

In the framework of BQE, our protocol sequentially updates the phase probability distribution with given measurement data.
To adaptively push spin-squeezed-state-based metrology working on its optimal point, we introduce a controllable auxiliary phase $\Phi_l$ for each iteration. 
Thus the $l$-th expectation of half-population difference can be rewritten as
\begin{equation}
    m^z_l(\phi;\Phi_l) = \tilde{\mathcal{A}}(\xi) \sin(\phi-\Phi_l)=\tilde{\mathcal{A}}(\xi) \sin(\tilde\phi_l),
\label{eq:singlelikelihood2}
\end{equation} 
where $\tilde{\mathcal{A}}(\xi)$ is a coefficient related to spin-squeezing parameter $\xi$ and we set $\tilde\phi_l=\phi-\Phi_l$ as the total phase for simplicity. 
The $l$-th likelihood function for the spin-squeezed state can be approximated as a Gaussian function versus the half-population difference,
\begin{equation}
    \mathcal{L}_l\left(m^z|\phi; \Phi_l\right)=\frac{1}{\sqrt{2\pi}\sigma_{l}^{m^z}} \exp\left[-\frac{(m^z-m^z_l)^2}{2(\sigma_{l}^{m^z})^2} \right],
\label{eq:likelihood}
\end{equation}
where $m^z=\frac{1}{2}(N_a - N_b)$ is the half-population difference between two modes and its uncertainty $\sigma_{l}^{m^z}=\tilde{\mathcal{A}}(\xi)\Delta\phi=\frac{\tilde {\mathcal{A}}(\xi) \xi}{\sqrt N}$ when $\tilde{\phi}_l\approx0$~\cite{MA201189}.
%

By using adaptive BQE, it is feasible to achieve high sensitivity~\cite{PhysRevX.9.021019,RN3,RN5,RevModPhys.89.035002,PhysRevA.85.030301,Cimini_2024} and strong noise robustness~\cite{PhysRevLett.118.100503,wang2017experimental,PhysRevA.65.043803,PhysRevLett.89.133602,Han:21} under limited measurements.
The auxiliary phase in our protocol is adaptively adjusted based on the previous measurement data. 
To ensure the spin-squeezed-state-based phase estimation working at its optimal point with steepest slope and highest precision~\cite{RevModPhys.89.035002,PhysRevResearch.7.L012064}, the $l$-th auxiliary phase is set as the $(l-1)$-th estimate: $\Phi_l=\phi_{est}^{l-1}$, see Fig.~\ref{BayesianPhaseEstimation}~(a). 
%
%
If there is no prior knowledge, the initial prior distribution can be set as a uniform distribution.
The posterior distribution is updated through Bayes’ inference: $p_l (\phi|m^z; \Phi_l)=\mathcal{N}_l\mathcal{L}_l(m^z|\phi; \Phi_l) p_{l-1} (\phi)$,
where $p_l$ and $p_{l-1}$ respectively denote the posterior and prior distributions, and $\mathcal{N}_l$ is the normalization factor. 
Then $\phi$ can be estimated as $\phi_{est}^{l} = \int \phi p_l (\phi|m^z; \Phi_l) d\phi$ with an uncertainty $\sigma^{\phi}_l = \sqrt{\int \phi^2 p_l (\phi|m^z; \Phi_l)d\phi - (\phi_{est}^{l})^2}$.
Compared to a spin-coherent state, the half-population difference for a spin-squeezed state can be approximated by a Gaussian distribution $\mathcal{N}(m^z_l, \sigma_{l}^{m^z})$ with a squeezed uncertainty $\sigma_{l}^{m^z}$.
Then, the $l$-th posterior can be given by the multiplication of a series of Gaussian functions~\cite{PhysRevResearch.7.L012064},
\begin{equation}
    p_l (\phi|m^z; \Phi_l)=\frac{1}{\sqrt{2\pi}\sigma_{l}^{\phi}} \exp\left[-\frac{(\phi-\phi_{est}^{l})^2}{2(\sigma_{l}^{\phi})^2}\right],
\label{eq:finalposterior}
\end{equation}
where, for sufficiently large $l$, the expectation $\phi_{est}^{l}$ approaches to the true value and the phase uncertainty can be given as $\sigma_{l}^{\phi}=\frac{\xi}{\sqrt{N}\sqrt{l}}$.

Below we analyze the performance of our adaptive BQE protocol with spin-squeezed states.
To ensure optimal metrological performance, our protocol measures spin-squeezed states at their optimal working points where quantum fluctuations is lowest.
As shown in Fig.~\ref{BayesianPhaseEstimation}~(b), the adaptive BQE can achieve the optimal phase measurement precision [Eq.~\eqref{optimalpoint}], showing an enhancement over the spin-coherent state with a factor of $\xi$.
A key advantage of our protocol is its inherent noise robustness. 
In most quantum phase estimation, there are two typical kinds of noises~\cite{PhysRevApplied.10.044033,PhysRevResearch.2.033078,Qiu_2022,PhysRevA.109.042412,10.21468/SciPostPhys.17.1.014,PhysRevResearch.7.L012064}: depolarization noise and phase noise.
Incorporating these noises, the likelihood function~\eqref{eq:singlelikelihood2} becomes 
\begin{equation}
    m^z_l(\phi;\Phi_l,\sigma_n) = (1-\tilde{p}_d)\tilde{\mathcal{A}}(\xi) \sin(\tilde\phi_l+\tilde\sigma_{n}).
\label{eq:noiselikelihood}
\end{equation}
Here the depolarization noise $\tilde p_d \sim |\mathcal{N} (0, p_d^{2})|$, which is a Gaussian distribution with $0\le \tilde{p}_d\le 1$, results in the contrast reduction. 
While the phase noise $\tilde {\sigma}_{n} \sim \mathcal{N} (0, {\sigma}_{n}^{2})$ leads to random errors. 
Thus the measurement precision depends on not only  quantum projection noise (QPN) but also these two noises. 

\begin{figure}[htbp]
\includegraphics[width=\linewidth]{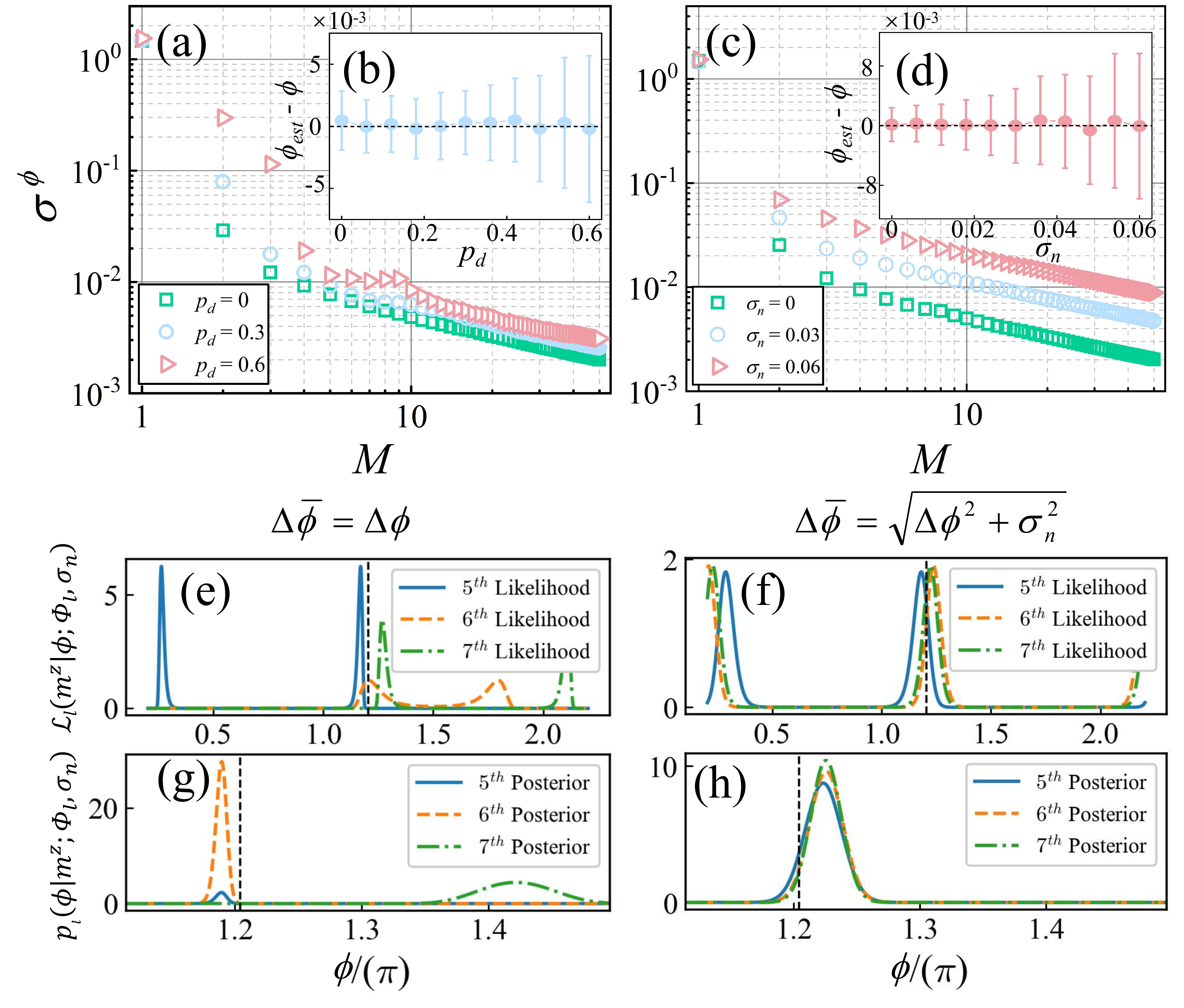}%
\caption{\label{Noiserobust} (color online).
Performances of Bayesian spin-squeezing-enhanced phase estimation under noises. 
The precision $\sigma^{\phi}$ versus the iteration times $M$ under (a) depolarization noise and (c) phase noise. 
The error $\phi_{est}-\phi$ versus $M$ under (b) depolarization noise and (d) phase noise. 
In our calculations, we choose $M$=50, $N$=200, and  $100$ repetitions for each point.
Under phase noise of strength $\sigma_{n} = 0.03$, the $5$-th to $7$-th likelihood functions are calculated using (e) the ideal uncertainty $\sigma_l^{m^z}$ and (f) the reshaped uncertainty $\bar{\sigma}_l^{{m}^z}$, and the corresponding $5$-th to $7$-th posterior distributions are shown in (g) and (h).}
\end{figure}

To quantitatively characterize the noise resilience of our protocol, we perform a statistical analysis by averaging over $100$ independent realizations for each noise strength. 
%
%
For depolarization noise, the precision decreases with the noise strength [see Figs.~\ref{Noiserobust}~(a)], and the errors $\phi_{est}-\phi$ (whose fluctuations are denoted by error bars) are slightly changed [see Figs.~\ref{Noiserobust}~(b)].
For phase noise, which can be regarded as a kind of technical noises, it is inherently present in experiments. 
If only QPN is considered, the likelihood with uncertainty $\sigma_l^{m^z} \propto \tilde{A}(\xi)\xi/\sqrt{N}$ suffers significant shape distortion under strong phase noise.
This distortion causes it to deviate from the ideal dual-Gaussian profile within a single phase cycle.
Consequently, the posterior uncertainty does not narrow monotonically throughout the iteration process, see Fig.~\ref{Noiserobust}~(e) and (g).
Thus the influence of phase noise should be taken into account~\cite{PhysRevApplied.22.044058,doi:10.1126/sciadv.adt3938}.
Since QPN and phase noise are independent, the uncertainty can be expressed as 
\begin{equation}
\begin{aligned}
    \bar{\sigma}_l^{{m}^z} &=\Delta{\bar\phi} \left| {{\partial m^z_l}/{\partial\tilde{\phi}_l}}\right|=\Delta{\bar\phi} \left| \tilde{\mathcal{A}}(\xi) \cos \tilde{\phi}_l \right|\\
    &=\sqrt{\Delta \phi^2+\sigma_{n}^2}\left| \tilde{\mathcal{A}}(\xi) \cos \tilde{\phi}_l \right|.
\label{eq:effectiveatomnumber}
\end{aligned}
\end{equation}
Therefore, the $l$-th likelihood function becomes 
\begin{equation}
    \mathcal{L}_l\left(m^z|\phi; \Phi_l,\sigma_n\right)=\frac{1}{\sqrt{2\pi}\bar{\sigma}_l^{m^z}} \exp\left[-\frac{(m^z-m^z_l)^2}{2(\bar{\sigma}_{l}^{m^z})^2} \right].
\label{eq:effectiveatomnumberlikelihood}
\end{equation}
This means that the phase uncertainty $\sigma^{\phi}$ sensitively depends on the phase noise $\sigma_{n}$.
However, due to the enhanced robustness of the reshaped Gaussian distributions during iteration, the error $\phi_{\text{est}}-\phi$ and its fluctuations are reduced even under strong phase noise~\cite{SM}, see Fig.~\ref{Noiserobust}(c) and (d).
Even when phase noise dominates, the noise-reshaped likelihood function maintains its stable shape [Fig.~\ref{Noiserobust}(f, h)], thereby enhancing estimation accuracy.

Furthermore, using the known relationship between them, the parameter $\gamma$ can be determined by measuring the accumulated phase $\phi$.
Usually, for a given interrogation time $T$, the phase-parameter relation can be expressed as $\phi=\mathcal{D}\gamma T^{\alpha}$ with $D$ and $\alpha$ related to the measurement protocol.
For atomic gravimetry~\cite{APeters_2001,PhysRevLett.117.203003} and atomic gyroscopes~\cite{PhysRevD.109.064010,RN9,PhysRevLett.134.143601}, $\alpha=2$.
While for atomic clocks~\cite{RevModPhys.87.637} and quantum magnetometry~\cite{PhysRevLett.104.133601,PhysRevX.5.031010,RevModPhys.90.025008}, $\alpha=1$.
Combining an interferometry sequence of optimized interrogation times $\{T_j\}$, the sub-SQL precision can be achieved while maintaining high dynamic range~\cite{qute202300329}.
Ideally, the $l$-th precision of $\gamma$ can be given as
\begin{equation}
    \Delta_l^{\gamma}=\frac{\xi}{C\mathcal{D}\sqrt{N} \sqrt{\sum_{j=1}^{l} (T_j^{\alpha})^2}},
\label{eq:parameterestimation}
\end{equation}
where $C$ is the contrast. 
The dynamic range, determined by the minimum interrogation time $T_1$~\cite{doi:10.1126/sciadv.adt3938}, remains unchanged even as the precision is gradually improved.

\textit{Bayesian gravimetry via spin-squeezed states. --} 
We take quantum gravimetry as the first example.
In frequentist estimation, it is challenging to simultaneously achieve high precision and high dynamic range. 
Typically, one should scan at least three fringes with different interrogation times to uniquely determine the gravity acceleration $g$~\cite{carraz2009compact,zhou2011measurement,abend2016atom,Lellouch03042022}.
In particular, unlike spin-coherent states, fringe-fitting for spin-squeezed states suffer significant precision degradation due to their narrow sub-SQL window, and consequently the achievable precision may be even worse~\cite{SM}.
However, our Bayesian protocol adaptively identifies the spin-squeezed state's optimal working point for each interrogation cycle, achieving enhanced precision beyond conventional approaches.  
Therefore, one can choose a suitable minimum interrogation time $T_1=T_{min}$ to ensure a high dynamic range, whereas exponentially increasing to $T_{max}$ to improve precision~\cite{2024arXiv241114944Z,PhysRevApplied.22.044058,doi:10.1126/sciadv.adt3938,PhysRevResearch.7.L012064}.

\begin{figure}[htbp]
\includegraphics[width=\linewidth]{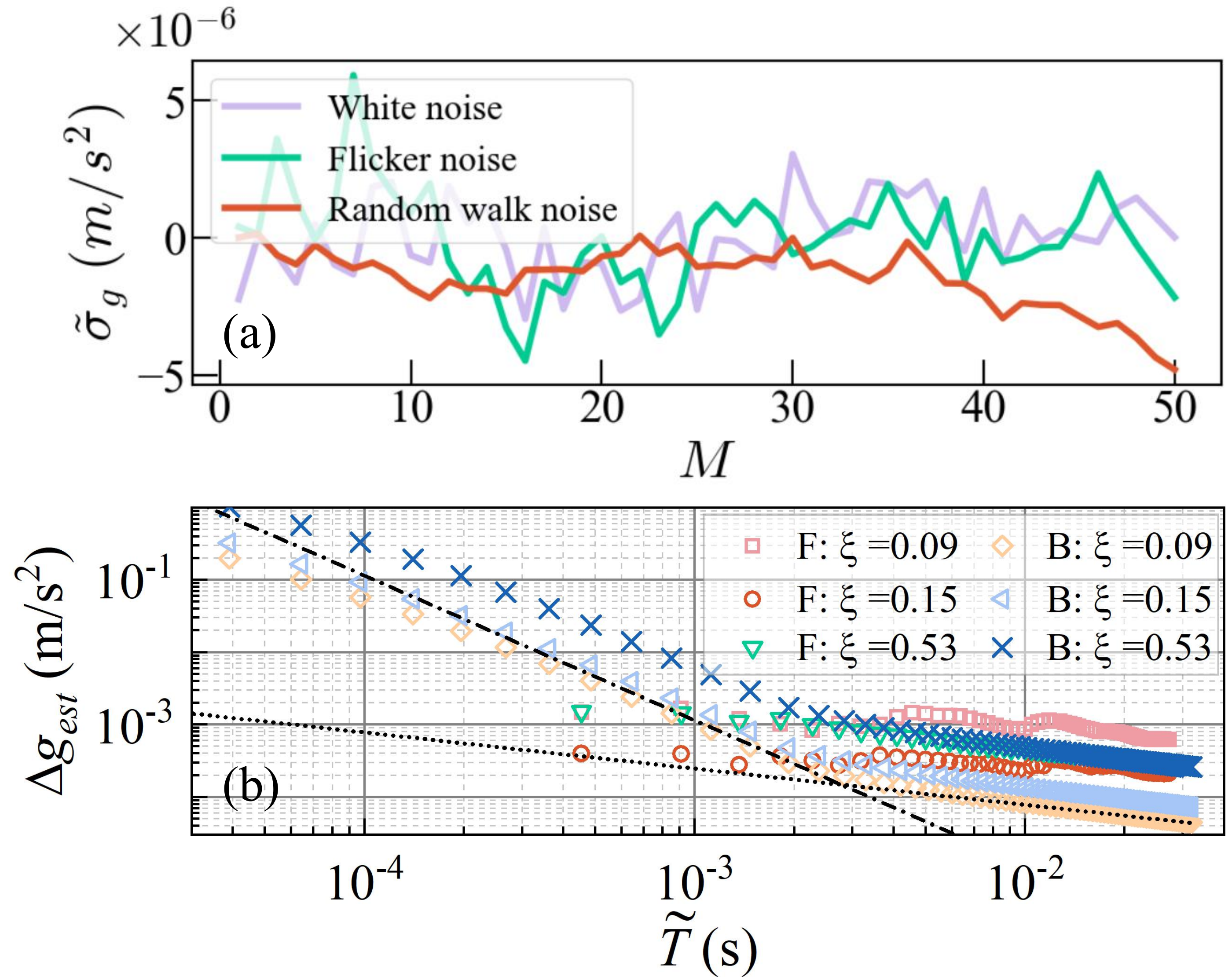}%
\caption{\label{GravimetryNoiserobust} (color online).
Bayesian gravimetry with spin-squeezed states under noises. 
(a) The distributions of white, flicker and random walk noises over $60$ trials. 
(b) The precision $\Delta g_{est}$ versus the total interrogation time $\tilde{T}$ for different spin-squeezed states under noises. 
Here, $B$ and $F$ denote Bayesian and Frequentist protocols, respectively. 
We set $N = 6000$, $T_{max} = 455~\mathrm{\mu s}$ and $C = 0.98$ based on Ref.~\cite{PhysRevX.15.011029}.}
\end{figure}

In gravimetry with cold $^{87}$Rb atoms, we have the phase $\phi = k_{\mathrm{eff}}gT^2$ with $k_{\mathrm{eff}}=1.61\times10^7$~rad/m and the gravity acceleration $g$.
According to Eq.~\eqref{eq:parameterestimation}, we have $\Delta_l^g=\frac{\xi}{C\sqrt{N} k_{\rm{eff}}\sqrt{\sum_{j=1}^{l} T_j^4}}$ in the absence of noises.
Using an exponentially increasing interrogation time sequence $\{T_l\}$~\cite{PhysRevResearch.7.L012064}, the precision scaling versus the total interrogation time $\tilde{T}=\sum_{j} T_j$ can be improved to $\Delta g \propto \tilde{T}^{-2}$.
While after $T_l$ reaches $T_{max}$ and remains fixed, the precision scaling reverts to $\Delta g \propto \tilde{T}^{-0.5}$.
There are three predominant types of phase noises in gravimetry: white noise $\sigma_w$, flicker noise $\sigma_f$, and random walk noise $\sigma_r$.
The white noise adheres to a Gaussian distribution $\tilde {\sigma}_{w} \sim \mathcal{N} (0, {\sigma}_{w}^{2})$. 
Flicker noise, which is generated by the Fourier-transformed Gaussian white noise under $1/\sqrt{f}$ spectral filter followed by inverse Fourier transform and variance normalization, results in a characteristic $1/\sqrt{f}$ power spectral density (PSD).
The random walk noise, which originates from cumulative integration of Gaussian white noise, produces a non-stationary process with a characteristic $1/f$ PSD.
We set $\sigma_w=\sigma_f=10 \sigma_r = 1\times10^{-6}~\mathrm{m/s^2}$ based on realistic noisy conditions~\cite{Lellouch03042022,rs16142634}, see Fig.~\ref{GravimetryNoiserobust}~(a).
As these noises appear simultaneously, the overall noise strength is $\sigma_g = \sqrt{\sigma_w^2+\sigma_f^2+\sigma_r^2}$.
Our protocol shows an order-of-magnitude improvement compared to conventional frequentist approaches~\cite{SM}, see Fig.~\ref{GravimetryNoiserobust}~(b).
The typical QPN is given by $\sigma_p=\xi/(C\sqrt{N}k_{\rm{eff}}T_{max}^2)\approx2\times10^{-3}~\mathrm{m/s^2}$.
Since the phase noise in current gravimeters is below $10^{-6}~\mathrm{m/s^2}$, its contribution to measurement imprecision is negligible, and thus our protocol is almost unaffected.

\begin{figure}[htbp]
\includegraphics[width=1.0\linewidth,scale=1.00]{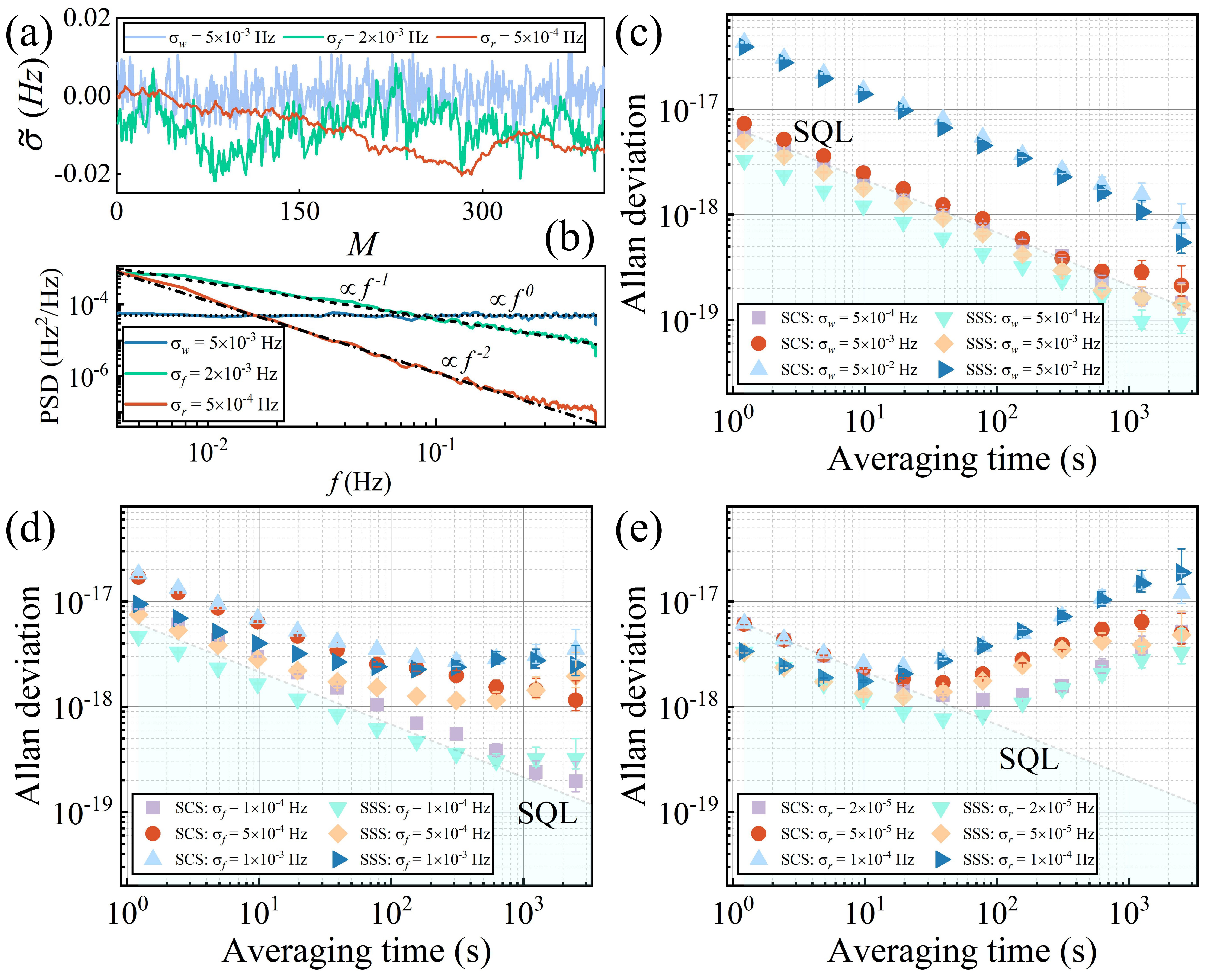}%
\caption{\label{Clocklocking.pdf} (color online).
Bayesian clock locking with spin-coherent state (SCS) and spin-squeezed state (SSS) under noises. 
(a) Distributions of white noise $(\sigma_w)$, flicker noise $(\sigma_f)$ and random walk noise $(\sigma_r)$ over $400$ trials. 
(b) PSD of $S_y(f)\propto f^{-\alpha}$ exhibiting different slopes, where $\alpha = \{0, 1, 2\}$ correspond to \{white, flicker, random walk\} noises. 
Allan deviations under (c) white noise, (d) flicker noise, and (e) random walk noise. 
Here, $T_{max} = 141$~ms, $N = 30000$, $C = 0.91$, and $\xi^2 = -5.1$ dB are chosen based on Ref.~\cite{yang2025clockprecisionstandardquantum}.}
\end{figure}

\textit{Bayesian clock locking with spin-squeezed states. --}
We take atomic clock locking as the second example, in which the phase $\phi = \omega T$ is proportional to the clock frequency $\omega$. 
With the Bayesian locking procedure~\cite{PhysRevApplied.22.044058}, we employ spin-squeezed state to improve the clock stability. 
The locking procedure is implemented via $12$ sequential interferometry measurements. 
The interrogation time $T_j$ increases exponentially for the first six steps (with a growth rate of $1.3$) and is then fixed as the maximum value $T_{\text{max}}$ for the remaining six steps.
The clock laser noise is characterized by a PSD of $S_y(f)= h_\beta f^{-\beta}$ with the exponent $\beta$ dependent on noise. 
The clock stability~\cite{2007Handbook,kielinski2025bayesianfrequencymetrologyoptimal} is characterized by the Allan deviation $\sigma_y^2(\tau)=h_{\beta}^2 \tau^{(\beta-1)}$ versus the averaging time $\tau$.
Here $\beta = \{0, 1, 2\}$ and $h_{\beta}^2=\{1/2\sigma_w^2, 2 \ln 2\sigma_f^2, 2\pi^2/3\sigma_r^2\}$ respectively correspond to white, flicker, and random walk noises, see Fig.~\ref{Clocklocking.pdf}~(a,~b).

Under white noise, the Allan deviation increases with the noise strength without changing its slope. 
As expected, the spin-squeezed state achieves a lower Allan deviation than the spin-coherent state.
However, this advantage gradually diminishes as the noise becomes dominant, see Fig.~\ref{Clocklocking.pdf}~(c).
Under flicker noise, the Allan deviations for both spin-squeezed and spin-coherent states converge to the same asymptotic value, which depends on the noise strength.
Nevertheless, the spin-squeezed state maintains superior short-term stability compared to the spin-coherent state, see Fig.~\ref{Clocklocking.pdf}~(d).
Under random walk noise, the Allan deviations for both spin-squeezed and spin-coherent states increase with the averaging time and ultimately converge to a common trend determined by the noise strength, see Fig.~\ref{Clocklocking.pdf}~(e).
Thus, under modest noise, the spin-squeezed state may provide enhanced precision over the spin-coherent state.

\textit{Conclusion and discussion. --} 
We present an adaptive Bayesian quantum estimation protocol that enables robust enhanced precisions with spin-squeezed states under noises. 
Our protocol integrates three crucial aspects: (i) adaptively locking to the optimal working points of spin-squeezed states, (ii) performing noise-resilient estimation with a reshaped likelihood function that accounts for phase noise, and (iii) implementing a correlated interferometry sequence to facilitate Bayesian iteration.
This protocol outperforms conventional methods in parameter estimation under realistic noises, achieving superior sensitivity and robustness without compromising the dynamic range.
Spin-squeezed states have been generated in various systems, such as Bose condensed atoms~\cite{riedel2010atom,gross2010nonlinear,esteve2008squeezing,doi:10.1126/science.1250147},
trapped ions~\cite{doi:10.1126/science.abi5226,doi:10.1126/science.aad9958}, and atoms in cavity~\cite{pedrozo2020entanglement,PRXQuantum.3.020308,RN17}.
%
%
Our work establishes a comprehensive framework for quantum-enhanced metrology that simultaneously addresses the key challenges of precision, dynamic range, and noise resilience in practical sensing.

\begin{acknowledgments}
This work is supported by the National Natural Science Foundation of China (Grants No.~12025509, No.~12475029 and No. 92476201), the National Key Research and Development Program of China (Grant No.~2022YFA1404104), and the Guangdong Provincial Quantum Science Strategic
Initiative (GDZX2305006 and GDZX2405002).
\end{acknowledgments}








\title{Supplementary Material for ``Practical robust Bayesian spin-squeezing-enhanced quantum sensing under noises''}










 

\begin{appendices}

\section{Appendix A: Spin-squeezing-enhanced phase estimation via one-axis twisting}
The fundamental process for quantum metrology employing spin-squeezed states comprises four principal stages: initial spin-squeezed state preparation, interrogation, readout, and estimation~\cite{RN13}.
Assuming that all $N$ particles are in state $\ket{a}$ initially, upon the application of a $\pi/2$ pulse rotating around the $J_y$-axis, they evolve into a spin-coherent state polarized along the $J_x$-axis with equal probabilities of being in states $\ket{a}$ and $\ket{b}$.
In this situation, the spin-coherent state can be written as
\begin{equation}
    \ket{\pi/2,0}_{SCS}=\otimes^{{N}}_{{l=1}}\left[\frac{1}{\sqrt{2}}\left(\ket{a}^{(l)}+\ket{b}^{(l)}\right)\right].
\label{eq:SCS}
\end{equation}

The most commonly used method to generate spin-squeezed states is one-axis twisting (OAT)~\cite{esteve2008squeezing,riedel2010atom,gross2010nonlinear,doi:10.1126/science.1250147,pedrozo2020entanglement,PRXQuantum.3.020308,PhysRevA.91.033625,doi:10.1126/science.aad9958,franke2023quantum,RN8,PhysRevLett.131.063401}.
Generally, the OAT Hamiltonian is 
\begin{equation}
    H_{OAT}=\chi J_z^2,
\end{equation}
where $\chi$ is OAT interaction strength. Initially from $\ket{\pi/2,0}_{SCS}$, the system state evolves into a spin-squeezed state through the OAT, i.e.,
\begin{equation}
    \ket{\Psi}_{SS}=e^{-i\chi t\hat{J}_z^2}\ket{\pi/2,0}_{SCS}.
\label{eq:SSS}
\end{equation}
Experimentally, a pulse with a specific phase can be applied to rotate the spin-squeezed state along $x$-axis with a rotation angle $\alpha$, i.e.,
\begin{equation}
    \ket{\Psi}_{SR}=e^{-i \alpha \hat J_x}\ket{\Psi}_{SS},
\label{eq:rotate angle}
\end{equation}
thereby aligning the direction of minimal fluctuation along the $J_y$-axis.

\begin{figure}[htbp]
\includegraphics[width=0.6\linewidth,scale=1.0]{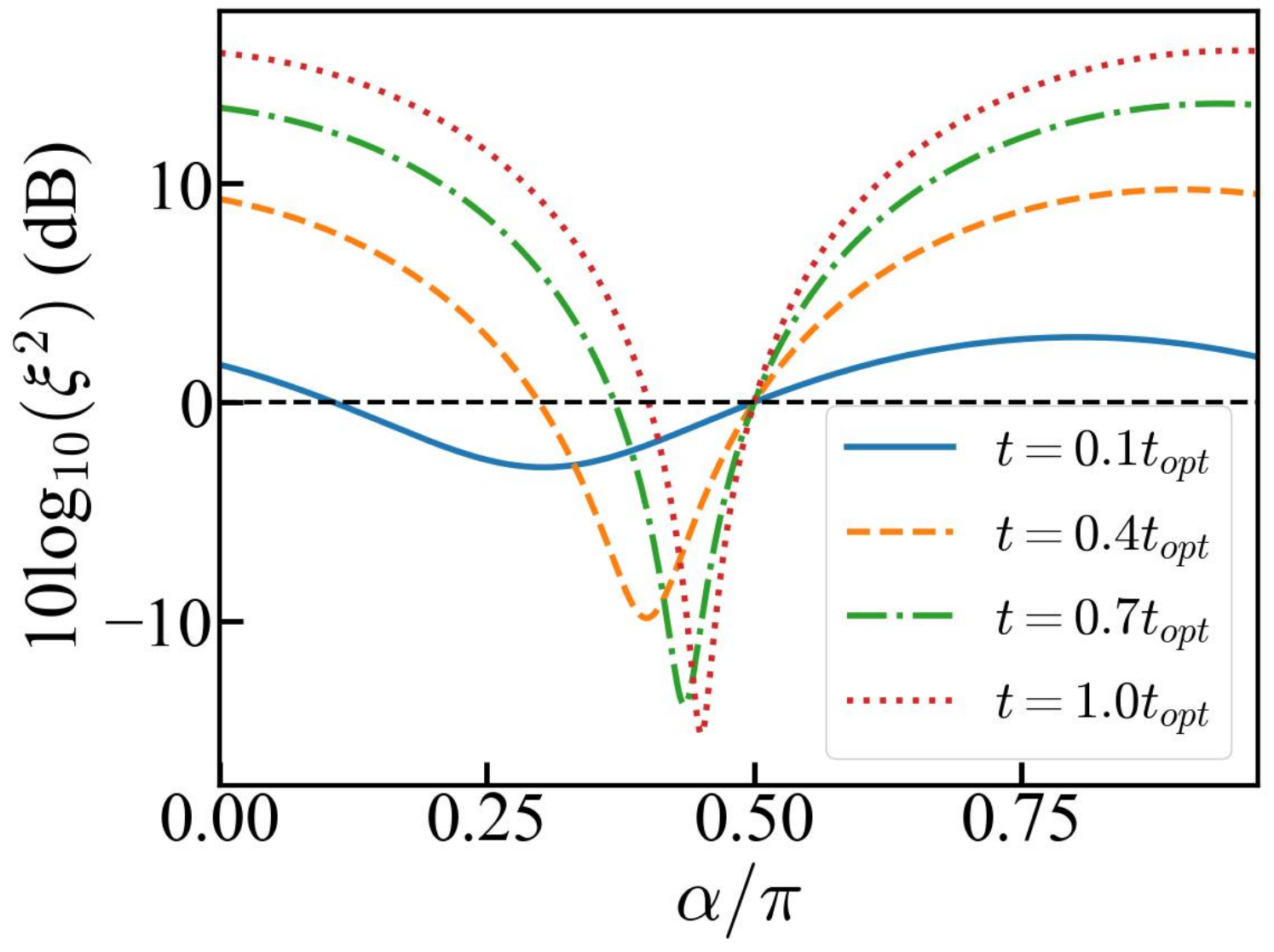}%
\caption{\label{Diffchit} \label{FigS1}
(color online)
The variation of the spin squeezing parameter versus the rotation angle for different one-axis twisting time $t$. Here, the optimal twisting time $t_{opt}=3^{1/3}N^{-2/3}/\chi$.}
\end{figure}

The squeezing parameter $\xi$ along the $J_z$-axis in the OAT-generated spin-squeezed state is a function of $\chi t$ and the rotation angle $\alpha$ and it reads~\cite{PhysRevA.47.5138,10.1063/5.0204102}
\begin{equation}
    \xi(\chi t, \alpha)=\frac{\sqrt{1+\frac{N-1}{4}\left[A-\sqrt{A^2+B^2}\cos(2(\alpha+\delta))\right]}}{\cos^{N-1}(\chi t)}
\label{eq:spinsqueezedparameter}
\end{equation}
where $A=1-\cos^{N-2}(2\chi t)$, $B=4\sin(\chi t)\cos^{N-2}(\chi t)$ and $\delta=\frac{1}{2}\arctan\left(\frac{B}{A}\right)$.
Thus, there exists an optimal rotation angle $\alpha_{opt}$ that minimizes the $\xi$, as shown in Fig.~\ref{FigS1}.
The variation of the squeezing parameter $10\log_{10}\xi^2$ on rotation angle $\alpha$ differs for different degrees of spin squeezing.
Before the optimal spin squeezing is achieved, a higher degree of spin squeezing results in a narrower range of rotation angles over which the squeezing parameter remains effective.
Consequently, achieving such squeezing necessitates precise experimental control over the rotation angle—a requirement that, fortunately, can be met with current technological capabilities.

Then the state undergoes a signal accumulation, encoding an estimated phase $\phi$.
%
Thus, the state after interrogation reads
\begin{equation}
    \ket{\Psi}_{SI}=e^{-i \phi \hat J_z}\ket{\Psi}_{SR}.
\label{eq:interrogation}
\end{equation}
%
%
Finally, a $\pi/2$ pulse is applied to coherently recombine the interference modes, leading to a final state as
\begin{equation}
    \ket{\Psi}_{SF}=e^{-i (\pi/2) \hat J_y}\ket{\Psi}_{SI}.
\label{eq:Final state}
\end{equation}

One can measure the half population-difference between the modes to obtain the information about $\phi$ as 
\begin{equation}
    \langle \hat{J}_z \rangle = -\frac{N}{2}\sin\phi\cos^{N-1}(\chi t),
    \label{eq:meanvalue}
\end{equation}
where $N$ is the total particle number, $\phi$ represents the phase to be measured, $\chi$ denotes the twisting strength, and $t$ is the twisting time.
We can also obtain~\cite{PhysRevA.110.042619} 
\begin{equation}
\begin{aligned}
    \langle\hat{J}_z^2\rangle &= \sin^2\phi\frac{N}{8}\left[N+1+(N-1)\cos^{N-2}(2\chi t)\right] \\
    &\quad + (\cos\theta\cos\phi)^2\frac{N}{8}\left[N+1-(N-1)\cos^{N-2}(2\chi t)\right] \\
    &\quad + (\sin\theta\cos\phi)^2\frac{N}{4} \\
    &\quad - \cos\theta\sin\theta\cos^2\phi\frac{N(N-1)}{2}\cos^{N-2}(\chi t)\sin(\chi t),
    \label{eq:squremeanvalue}
\end{aligned}
\end{equation}
where $\theta$ denotes the rotation angle for the OAT-generated spin-squeezed state.
According to the error propagation formula, we have
\begin{equation}
    \Delta \phi = \frac{\Delta \hat{J}_z}{\lvert\frac{\partial\langle\hat{J}_z\rangle}{\partial\phi}\rvert}
   =\frac{\sqrt{\langle\hat{J}_z^2\rangle-\langle \hat{J}_z \rangle^2}}{\lvert\frac{\partial\langle\hat{J}_z\rangle}{\partial\phi}\rvert}  =\sqrt{\mathcal{A}\tan^2\phi+\mathcal{B}}. 
\label{eq:errorpropagation}
\end{equation}
Here, 
\begin{equation}
    \mathcal{A}=\frac{N+1+(N-1)\cos^{N-2}{2\chi t}}{2N\cos^{2(N-1)}{\chi t}}-1,
\end{equation}
and 
\begin{equation}
    \mathcal{B}=\xi^2/N. 
\end{equation}
The minimal phase fluctuation 
\begin{equation}
    \Delta \phi = \frac{\Delta \hat{J}_z}{\lvert\frac{\partial\langle\hat{J}_z\rangle}{\partial\phi}\rvert}={\xi/\sqrt{N}}, 
\end{equation}
for spin-squeezed state can obtain when $\phi = 0$. 
In particular, the optimal phase precision $\phi\propto N^{-5/6}$ can be obtained when $\chi t_{opt}=3^{1/3}N^{-2/3}$, which beats the standard quantum limit (SQL).

However, for $\phi>0$, the phase fluctuation quickly deviates from ${\xi/\sqrt{N}}$. 
That is, greater quantum projection noise will be introduced if we do not perform measurement at $\phi = 0$, thereby compromising the precision.
Developing an adaptive phase measurement protocol centered at $\phi=0$ to suppress quantum projection noise and achieve optimal metrological precision presents a significant technical challenge.

\section{Appendix B: Bayesian phase estimation with OAT-generated spin-squeezed states}

\begin{figure}[htbp]
\includegraphics[width=\linewidth,scale=1.0]{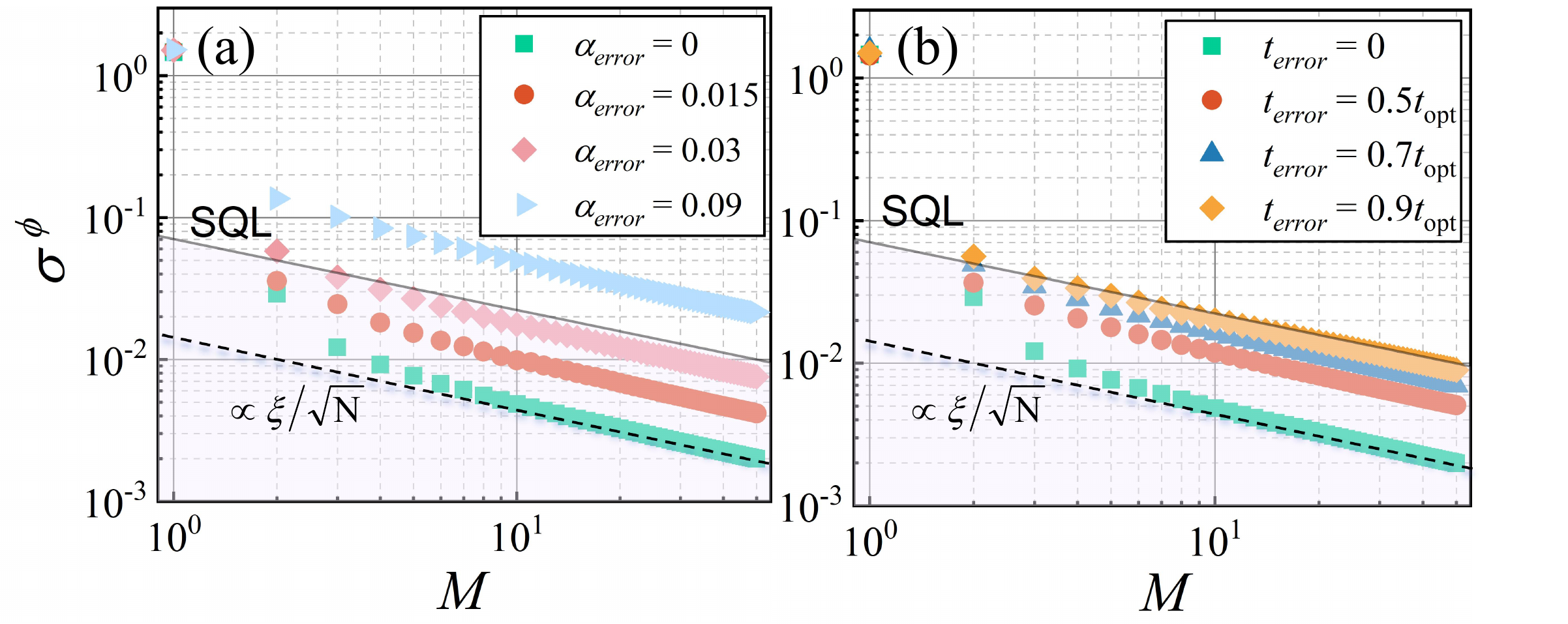}%
\caption{\label{terror} 
(color online) Influences of imperfect operations on the measurement precision with OAT-generated spin-squeezed states. (a) The precision of Bayesian phase estimation using optimal OAT dynamics with different turn angle error $\alpha_{error}$. (b) The precision of Bayesian phase estimation with optimal turn angle of $\chi t_{opt}$ under twisting time error $t$. Here, $M=50$ and $N=200$ are chosen.}
\end{figure}

For spin-squeezed states, we can perform population-difference detection as Eq.~\eqref{eq:meanvalue} and subsequent construction of likelihood function in the framework of Bayesian estimation.
This process is iteratively repeated to enable sustained Bayesian updating, thereby progressively refining the posterior phase estimation.
For OAT-generated spin-squeezed state, an additional turn $e^{-i\alpha\hat J_x}$ should be added to achieve the optimal spin-squeezed state for measurement. If the turn angle $\alpha$ is away from the optimal $\alpha_{opt}$ (as shown in Fig.~\ref{FigS1}), the angle error $\alpha_{error}=\alpha-\alpha_{opt}$ would induce a deterioration of measurement precision, see Fig.~\ref{terror}~(a).
However, the turn angle can be controlled precisely~\cite{pedrozo2020entanglement,franke2023quantum,eckner2023realizing}, ensuring the quantum-enhanced precision in experiments.

The properties of spin-squeezed states are highly sensitive to the experimental parameter $\chi t$, and discrepancies often arise between its empirically determined value and the theoretical optimal one, $\chi t_{opt}$.
If a measurement is performed using the optimal rotation angle for $\chi t_{opt}$ when the actual squeezing parameter differs from this value, the resulting precision will also be degraded and fail to reach the fundamental limit. As illustrated in Fig.~\ref{terror}~(b), the measurement precision exhibits an inverse dependence on the deviation of $t$ from $t_{opt}$.
Notably, even with such imperfect control, our measurement protocol still enables a precision that surpasses the SQL. 

\begin{figure}[htbp]
\includegraphics[width=0.8\linewidth,scale=1.00]{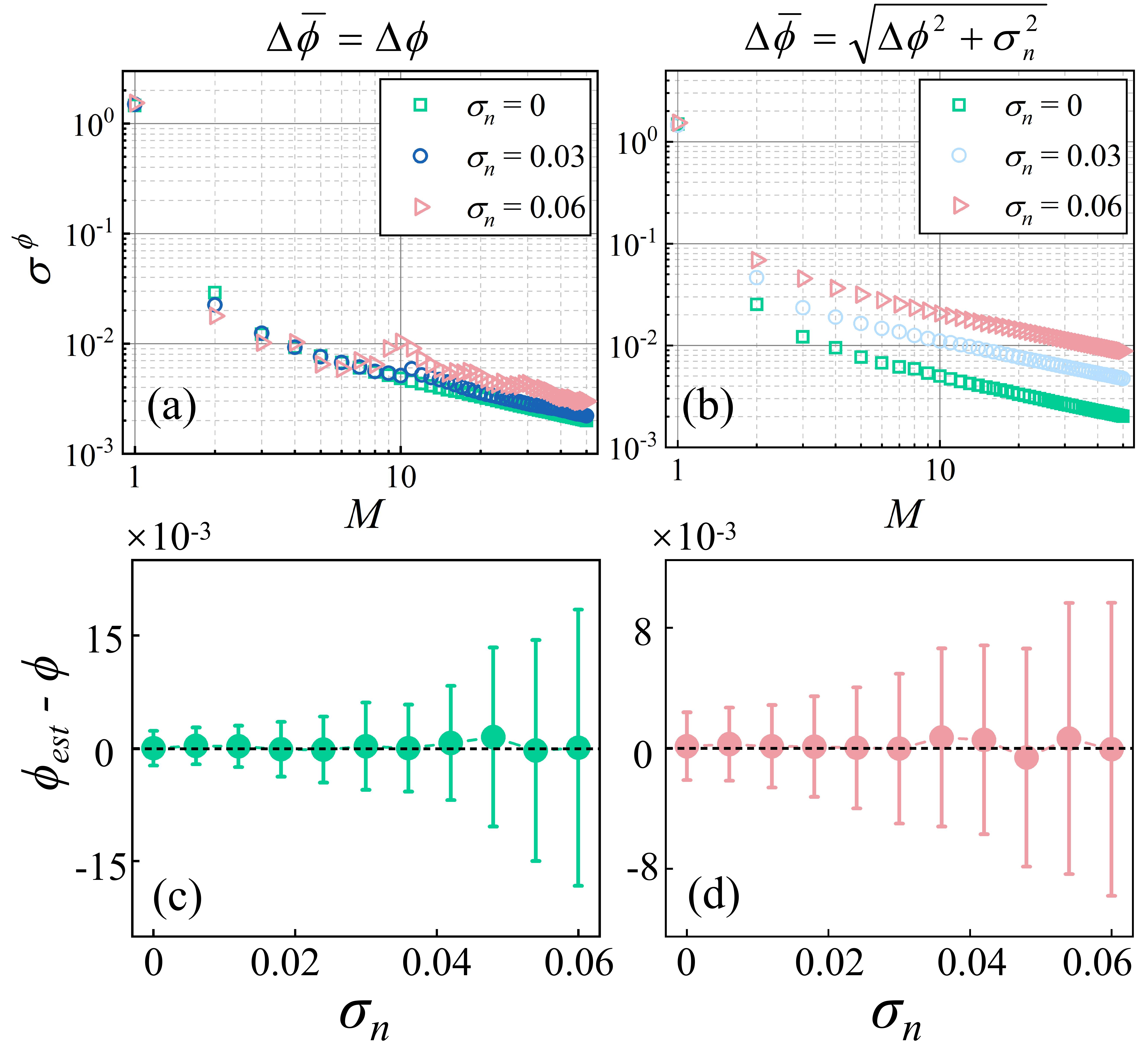}%
\caption{\label{Noiserobusteffatom} (color online)
Performances of Bayesian phase estimation in the presence of phase noises.  
Under different phase noise of strength, precision $\sigma^{\phi}$ are calculated using (a) the ideal uncertainty $\Delta {\phi}$ and (b) the reshaped uncertainty $\Delta {\bar{\phi}}$, and the corresponding error $\phi_{est}-\phi$ are shown in (c) and (d).
For each noise intensity, $100$ repetitions of Bayesian iteration are performed here. The black dashed line indicates the true phase value. In our numerical simulation, $M$=50 and $N$=200 are chosen.}
\end{figure}

\section{Appendix C: Influences of phase noise in Bayesian phase estimation}

The uncertainty in quantum phase estimation originates from two principal sources: quantum projection noise and phase noise. In practice, both contribute to the phase measurement fluctuations. 
Since the quantum projection noise and phase noise are independent, the fluctuation of $\phi$ in the likelihood can be rewritten as 
\begin{equation}
    \Delta \bar \phi = \sqrt{\Delta \phi^2+\sigma_n^2}.
\end{equation}
In the absence of $\sigma_n$, only the quantum projection noise contributes and the phase uncertainty in the likelihood reduces to $\Delta \bar \phi=  \Delta \phi$.

Incorporating the effects of phase noise into the likelihood function leads to broader posterior distribution widths, which leads to enhanced fluctuations in phase estimation, see Fig.~\ref{Noiserobusteffatom}~(a) and Fig.~\ref{Noiserobusteffatom}~(b).
However, when phase noise dominates, incorporating its effect on $\Delta \bar\phi$ leads to reduced fluctuations in the accuracy, see Fig.~\ref{Noiserobusteffatom}~(c) and Fig.~\ref{Noiserobusteffatom}~(d).
This is because incorporating phase noise into the likelihood function prevents significant distortion of the likelihood function under phase-noise-dominated conditions.
A stable morphology of the likelihood function is essential for ensuring the effective progression of Bayesian iterative procedures, which in turn reduces fluctuations in the phase estimate.
Therefore, under noisy conditions, the influence of phase noise must be explicitly incorporated into the construction of the likelihood function.

\section{Appendix D: Precision scalings of Bayesian atomic gravimetry}

In this section, we show how to implement the Bayesian atomic gravimetry with more details and analyze its precision scalings.
Conventional fringe-fitting methods require measuring at least three fringes with different interrogation times to determine the unambiguous $g$. However this determination process requests a lot of measurement times in experiments. In contrast, Bayesian quantum estimation may combine measurements taken with different interrogation times by updating the probability distribution with Bayes’ theorem. It is not necessary to perform the pre-estimation process. 
Within the Bayesian framework, we can utilize an increasing time series to achieve high dynamic range gravity estimation while maintaining precision. 

For atomic gravimetry, during the interrogation process, an auxiliary phase can be introduced, which can be achieved by controlling the linear chirp rate of the Raman laser.
In Bayesian quantum estimation (BQE), an auxiliary phase can be adaptively configured to achieve optimal measurement precision for spin-squeezed states.
Here, we can introduce an auxiliary parameter $g_c = 2\pi \alpha_c/k_{\rm{eff}}$, where $\alpha_c$ is the linear chirp rate of the Raman laser. 
By setting $g_c^l = g_{est}^{(l-1)}$, the measurement protocol dynamically converges, thereby enabling optimal precision for gravity measurements using spin-squeezed states.
For gravity measurement, the uncertainty with spin-squeezed state (only taking quantum projection noise into account) is 
\begin{equation}
    \Delta g = \xi/(\sqrt{N}k_{\mathrm{eff}}T^2), 
\end{equation}
which is inversely proportional to the interrogation time.
\begin{figure}[htbp]
\includegraphics[width=0.88\linewidth,scale=1.00]{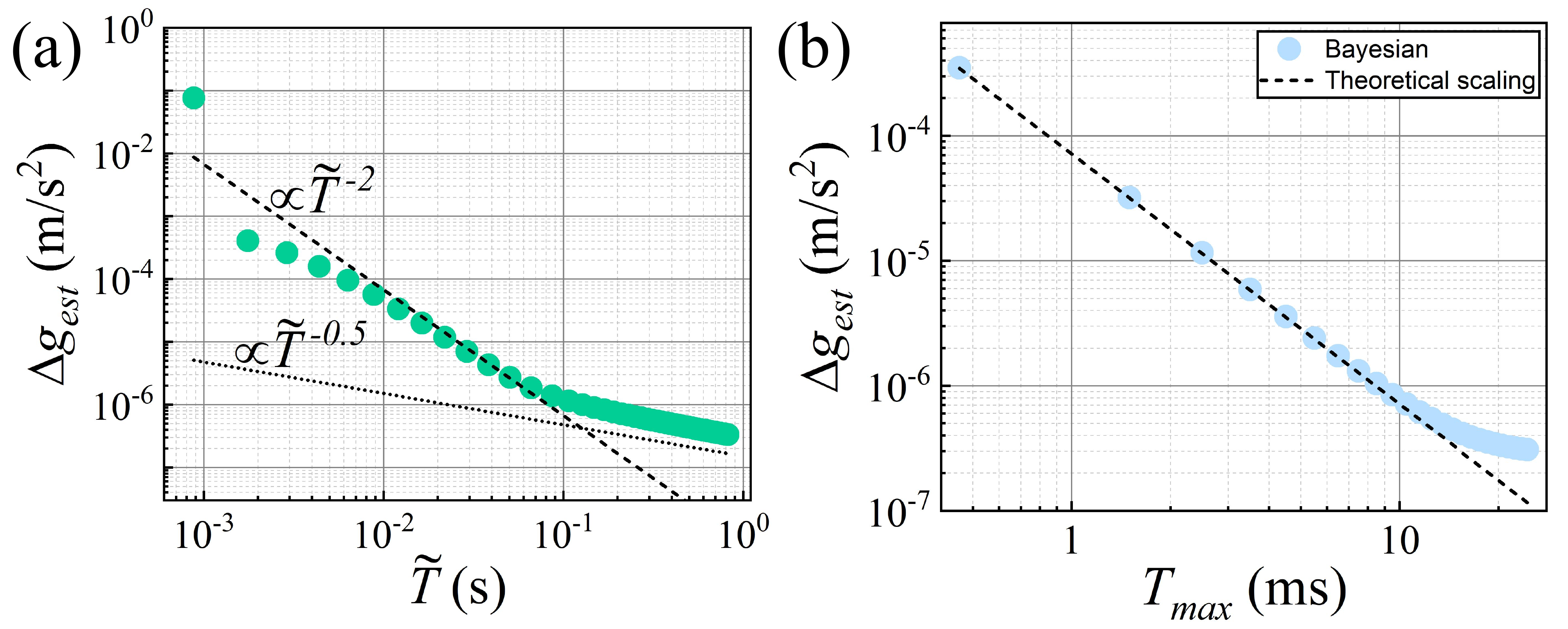}%
\caption{\label{GravimetrychangeT} (color online)
Precision scalings of Bayesian atomic gravimetry in the presence of noises. (a) The precision $\Delta g_{est}$ versus total interrogation time $\tilde{T}=\sum_l T_l$ with $T_l \le T_{max} = 20$~ms in the condition of $\sigma_w=\sigma_f=10 \sigma_r = 1\times10^{-6} ~\mathrm{m/s^2}$. (b) The influence of noise on precision across different maximum interrogation time $T_{max}$. Theoretical scaling represents the precision versus maximum interrogation time for $50$ iterations under noise-free conditions.
The BQE procedure is implemented via $M$ sequential interferometry measurements with the time sequence as Eq.~\eqref{eq:Ti}.
When $T_{max}$ increases, the quantum projection noise becomes smaller and the phase noise degrades the measurement precision. 
In our numerical simulation, $M=50$, $C=0.98$ and $N=6000$ are chosen based on Ref.~\cite{PhysRevX.15.011029}.}
\end{figure}

We analyze the impact of an increasing time series on the precision scaling of gravity measurement.
Generally, in the absence of prior information, the first prior is set to a uniform distribution function.
The first posterior then equals the first likelihood function, and their uncertainty is the same.
\begin{equation}
    \Delta g_{est}^{(1)} = \xi/(\sqrt{N}k_{\rm{eff}}T_1^2).
\end{equation}
%
By multiply the Gaussian-shaped likelihood step by step, we can obtain the $l$-th standard deviation as
\begin{equation}
    \Delta g_{est}^{(l)}=\frac{\xi}{\sqrt{N} k_{\rm{eff}}\sqrt{\sum_{j=1}^{l} T_j^4}}.
\label{eq:sigman}
\end{equation}
Therefore, the $l$-th posterior function can be analytically calculated as 
\begin{equation}
\ p_l (g|m^z; g_c^{(l)})=\mathcal{N}\prod_{j=1}^{l}\mathcal{L}_j=\mathcal{N}\prod_{j=1}^{l}\left[\frac{1}{\sqrt{2\pi}\sigma_j}\exp(-\frac{(g-\mu_j)^2}{2\sigma_j^2})\right],
\label{eq:p_n}
\end{equation}
where $\mathcal{N}$ is the normalization factor and the standard deviation for measuring $g$ is 
\begin{equation}\label{Delta_g}
    \Delta g_{est}^{(l)}=\frac{\xi}{\sqrt{N} k_{\rm{eff}}\sqrt{\sum_{j=1}^{l} T_j^4}}.
\end{equation}

Here, we set $\tilde{T}=\sum_{j} T_j$ as the total interrogation time.
%
If the interrogation time for each measurement is the same, e.g., $T_j=T_{max}$, the total interrogation time in the $l$-th step is $\tilde{T}_l = l T_{max}$. 
In this case we can obtain
\begin{equation}
    \Delta g_{est}^{(l)} = \frac{\xi}{\sqrt{N} k_{\rm{eff}}T_{max}^{3/2}\sqrt{\tilde{T}_l}} \propto \frac{\xi}{\tilde{T}_l^{0.5}},
\label{eq:SQL}
\end{equation} 
which is the scaling of SQL. 

The precision scaling depends on the form of the interrogation time series, and it can be further improved by choosing a suitable one. 
We consider the case in which $T_{l}$ grows exponentially with increment ratio $a$. Thus the $l$-th interrogation time $T_{l}=T_{1}a^{l-1}$ and the total interrogation time in the $l$-th step is $\tilde{T}_l=T_{1}(1-a^{l})/(1-a)\approx T_{l}a/(a-1)$ if $l\gg1$. Substituting $T_{l}$ and $\tilde{T}_l$ into Eq. \eqref{eq:sigman}, we can obtain
\begin{equation}
\begin{aligned}
\Delta g_{est}^{(l)}&=\frac{\xi}{\sqrt{N} k_{\rm{eff}}\sqrt{\sum_{j=1}^{l} T_j^4}}\\
&=\frac{\xi}{\sqrt{N} k_{\rm{eff}}T_{l}^2\sqrt{\sum_{j=0}^{l-1} a^{-4j}}}\\
&\approx\frac{\sqrt{a^{4}-1}}{\sqrt{N} k_{\rm{eff}}\tilde{T}_l^2(1-a)^2}\propto \frac{\xi}{\tilde{T}_l^{2}}.
\label{eq:sigmanunlimited}
\end{aligned}
\end{equation}
Therefore, we finally obtain the scaling of the standard deviation of measuring $g$ with exponential increasing time series is $\Delta g \propto \tilde{T}_l^{-2}$. 

However, in practise, the length of the atomic gravimeter cavity is finite, thus $T_{l}$ cannot be increased unlimitedly.
We denote the available maximum interrogation time as $T_{max}$ and take the exponential increasing time series as an example. If $T_{max}$ exists, $T_l$ first grows exponentially from $T_{1}$ to $T_{max}$ and continues to stay at $T_{max}$. Assuming one needs $M_a$ steps to increase from $T_1$ to $T_{max}$ and once $T_l$ reaches $T_{max}$, it keeps fixed at $T_{max}$ for the remaining $M-M_a$ steps, i.e., 
\begin{equation}
T_l = 
\begin{cases}
   T_{max}/a^{M_a-l}, & 1 \le l <M_a,\\
   T_{max}, &  M_a \le l \le M.
\end{cases}
\label{eq:Ti}
\end{equation}
Thus if $M_a < l \le M$, the total interrogation time in the $l$-th step is $\tilde{T}_l \approx T_{max}a/(a-1) + (l - M_a)T_{max} \approx (l - M_a)T_{max}$ when $l-M_a \gg {a}/{a-1}$. By substituting $T_{l}$ and $\tilde{T}_l$ into Eq.~\eqref{eq:sigman}, we can get 
\begin{equation}
\begin{aligned}
\Delta g_{est}^{(l)}&=\frac{\xi}{\sqrt{N} k_{\rm{eff}}\sqrt{\sum_{j=1}^{M_a} T_j^4+(l-M_a)T_{max}^4}}\\
&\approx\frac{\xi}{\sqrt{N} k_{\rm{eff}}\sqrt{(l-M_a)T_{max}^4}}\\
&=\frac{\xi}{\sqrt{N} k_{\rm{eff}}T_{max}^2\sqrt{\tilde{T}_l/T_{max}}}\\
&=\frac{\xi}{\sqrt{N} k_{\rm{eff}}T_{max}^{3/2}\sqrt{\tilde{T}_l}} \propto \frac{\xi}{\tilde{T}_l^{0.5}},
\label{eq:sigmanlimted}
\end{aligned}
\end{equation}
where the precision scaling eventually converges to the SQL $\tilde{T}_l^{-0.5}$ as Eq.~\eqref{eq:SQL} when the iteration times of using $T_{max}$ is large enough. Similarly, this result of Eq.~\eqref{eq:sigmanlimted} is also valid for using linear increasing scheme when the iteration times of using $T_{max}$ at the final stage is large enough. 

However in practice, the precision of the measurement will be affected by noise.
The influence is related to the magnitude of the quantum projection noise, which in turn is inversely proportional to the interrogation time.
In our protocol, the adoption of exponentially increasing interrogation time leads to distinct noise regimes: during shorter interrogation intervals, quantum projection noise remains the dominant factor, thereby rendering the measurement precision virtually unaffected by noises.
Despite the presence of $10^{-6}~\mathrm{m/s^2}$ level noise contributions, the $\propto \tilde{T}^{-2}$ scaling of measurement precision remains preserved with respect to the total interrogation time.
However, with a large maximum interrogation time, e.g. $T_{max}=20$ ms, in which both quantum projection noise and phase noise reach comparable magnitudes on the order of $10^{-6}~\mathrm{m/s^2}$ level, a significant reduction of measurement precision occurs.
Meanwhile, the time scale would also be changed to $\propto \tilde{T}^{-0.5}$, see Fig.~\ref{GravimetrychangeT}~(a).
Thus we also investigate the variation of measurement precision versus different maximum interrogation time $T_{max}$.
While extended interrogation time suppresses quantum projection noise, it simultaneously exacerbates the degradation of measurement precision caused by technical noise.
Consequently, the measurement precision deviates progressively from the theoretically optimal scaling,
as shown in Fig.~\ref{GravimetrychangeT}~(b).

\begin{figure}[htbp]
\includegraphics[width=0.88\linewidth,scale=1.00]{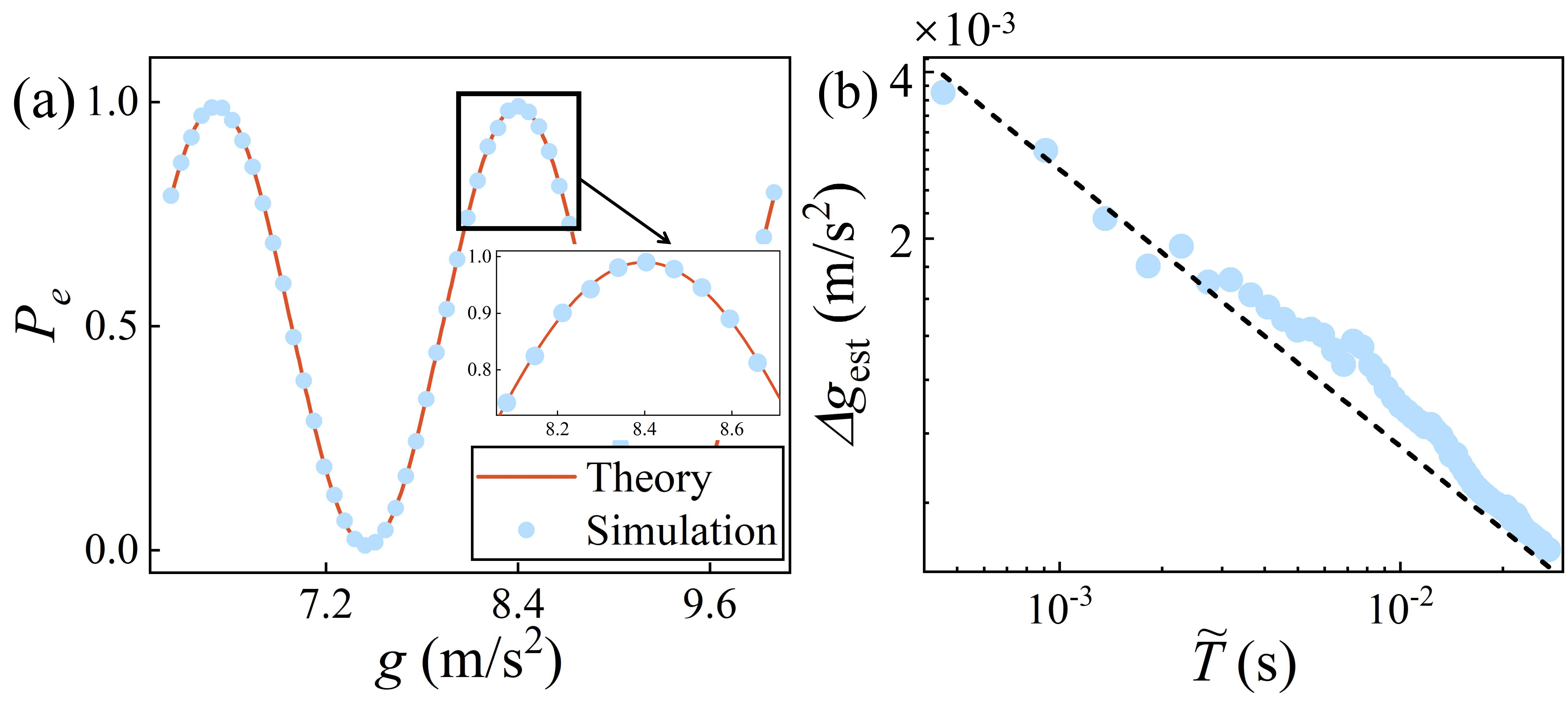}%
\caption{\label{CSSFrequencyfitting} (color online)
Performances of Frequentist estimation methodology (fringe fitting) for gravimetry with spin-coherent state in the absence of technical noises.  (a) The probability of excited state $(P_e)$ as a function of $g$ for spin-coherent state. The scatter points are the simulation results using fringe fitting and the solid red line is the theoretical one for comparison. (b) The measurement precision of $g$ versus total interrogation time $\tilde T= l T_{max}$ ($l=1,2,...,M$). The scatter points are the simulation results with fringe fitting and the black dashed line is the SQL ($\propto 1/\tilde T^{0.5}$). In our numerical simulation, $N = 6000$, $T_{max} = 455~\mathrm{\mu s}$, $C = 0.98$ are chosen.}
\end{figure}

\section{Appendix E: Fringe fitting method}

\begin{figure}[htbp]
\includegraphics[width=0.88\linewidth,scale=1.00]{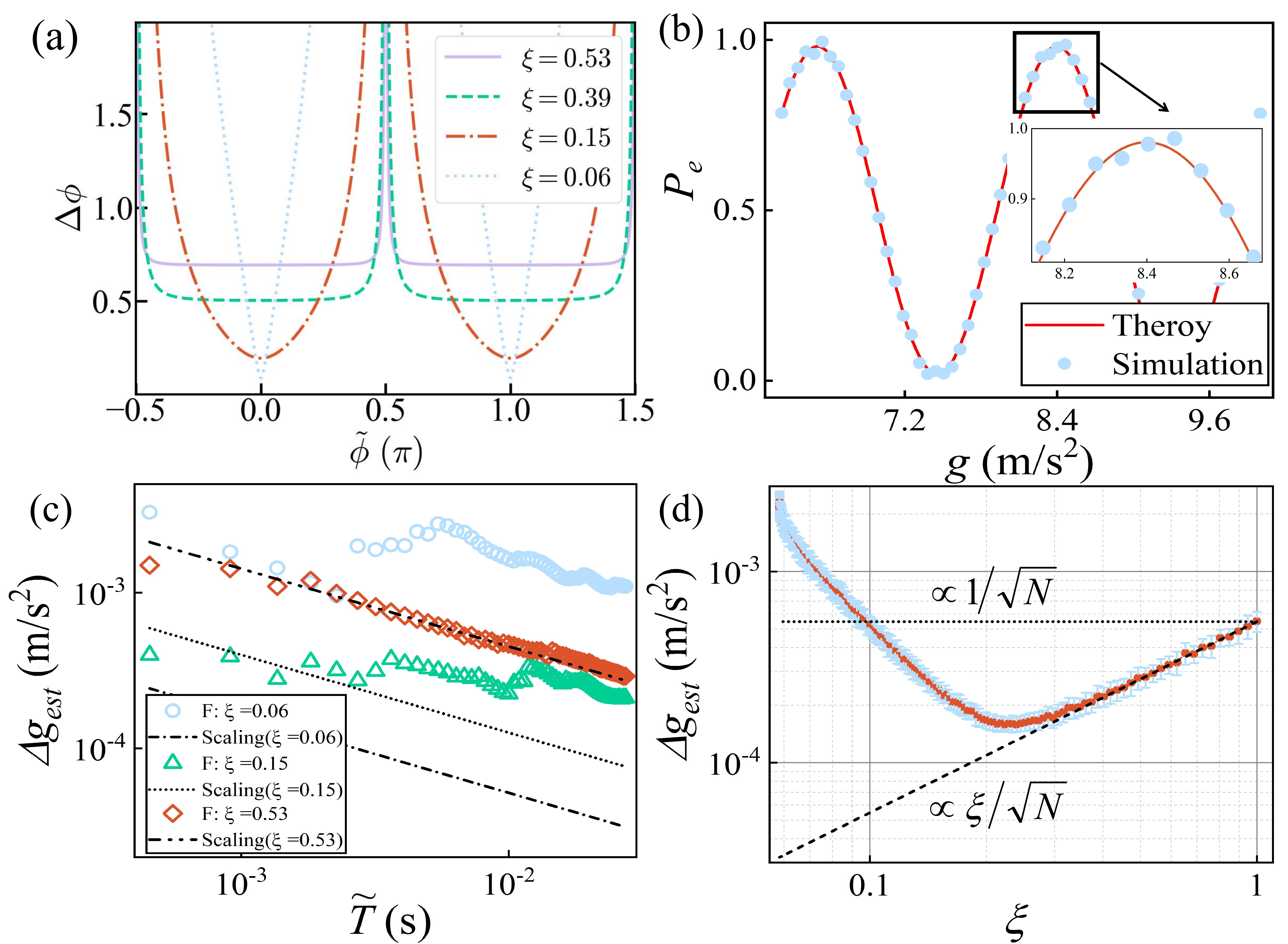}%
\caption{\label{Frequencyfitting} (color online)
Performances of Frequentist estimation methodology (fringe fitting) for spin-squeezed-state-based gravimetry in the absence of technical noises. (a) Phase measurement precision $\Delta \phi$ versus total phase $\tilde{\phi}$ for different degrees of spin squeezing according to Eq.~\eqref{eq:errorpropagation}. (b) The probability of excited state $(P_e)$ as a function of $g$ for $\xi=0.06$. The scatter points are the simulation results using fringe fitting and the solid red line is the theoretical one for comparison. (c) The measurement precision of $g$ versus total interrogation time $\tilde{T}=l T_{max}$ ($l=1,2,...,M$) for different degrees of spin squeezing. The scatter points are the simulation results with fringe fitting and the lines are the corresponding theoretical precision scalings. (d) The measurement precision of $g$ via the fringe fitting method versus the spin squeezing parameter $\xi$. The red dotted line represents the mean values of the fitting precision, and the error bars indicate the corresponding fluctuations, with $50$ repeated simulations conducted for each $\xi$. The dotted line represents the SQL, and the dashed line indicates the ultimate precision bound using spin-squeezed state. In our numerical simulation, $N = 6000$, $T_{max} = 455~\mathrm{\mu s}$, $C = 0.98$ are chosen based on Ref.~\cite{PhysRevX.15.011029}.}
\end{figure}

The quantum projection noise of a spin-squeezed state is strongly phase-dependent, with measurements away from $\phi = 0$ introducing significant uncertainty. 
This dependence intensifies with the squeezing strength, thereby narrowing the phase window of sub-SQL sensitivity [see Fig.~\ref{Frequencyfitting}~(a)]. 
Here using atomic gravimetry as an example, we demonstrate how conventional fringe-fitting methods fail to harness the full potential of spin-squeezed states, leading to suboptimal performance.

First, the measurement precision for a spin-coherent state is independent of the estimated phase. 
It maintains a uniform precision of $1/(\sqrt{N}k_{\mathrm{eff}}T^2)$ across the entire phase space, as its quantum projection noise remains constant, see Fig.~\ref{CSSFrequencyfitting}~(a). Consequently, under noiseless conditions, the precision of conventional fringe-fitting asymptotically converges to the SQL, see Fig.~\ref{CSSFrequencyfitting}~(b).

However, operating with spin-squeezed states face a fundamental challenge: the optimal sensitivity of these states is confined to a narrow phase window around $\phi = 0$, while their quantum projection noise increases dramatically elsewhere, see Fig.~\ref{Frequencyfitting}~(a). 
This strong phase-dependence, which intensifies with the squeezing strength (i.e., with decreasing spin-squeezing parameter $\xi$), renders standard full-fringe fitting protocols highly suboptimal.

Unlike the phase-independent precision of spin-coherent states — which is uniformly $1/(\sqrt{N}k_{\mathrm{eff}}T^2)$ and asymptotically reaches the SQL —fringe-fitting with squeezed states yields a precision that is both significantly worse than the theoretical optimum and non-monotonically dependent on $\xi$. 
This is because stronger squeezing amplifies the noise at non-optimal phase points, causing the precision to deviate rapidly from its expected value. Consequently, there exists a specific, optimal squeezing strength that maximizes the performance of this conventional method, as precision does not improve monotonically with $\xi$, see Fig.~\ref{Frequencyfitting}~(c) and (d). For example, the precision at $\xi=0.15$ surpasses that of both weaker ($\xi=0.53$) and stronger ($\xi=0.06$) squeezing.
As shown in Fig.~\ref{Frequencyfitting}~(d), the measurement precision of fringe-fitting for highly spin-squeezed state (with small $\xi$) is even worse than the SQL. 
Thus, without an adaptive measurement strategy, the practical benefit of spin-squeezing is limited, and simply increasing the squeezing strength can actually degrade measurement performance under standard fringe-fitting protocols.

It should be mentioned that, the results presented thus far assume ideal, noiseless conditions. Under realistic conditions involving technical noise, the measurement precision of the fringe-fitting method would be further degraded, rendering it impractical for use with spin-squeezed states. 

\end{appendices}

%

\end{document}